\documentclass[manuscript, authorversion]{acmart}
\renewcommand\footnotetextcopyrightpermission[1]{}
\usepackage{float}
\usepackage{booktabs}
\usepackage{longtable}
\usepackage{url}
\usepackage{soul}

\renewcommand{\arraystretch}{1.2}
\AtBeginDocument{%
  }

\setcopyright{acmlicensed}
\copyrightyear{2025}
\acmYear{2025}
\acmDOI{XXXXXXX.XXXXXXX}
\acmConference['CHI 26]{The ACM CHI Conference on Human Factors in Computing Systems}{April 13–-17, 2026}{Barcelona, Spain}
\acmISBN{978-1-4503-XXXX-X/2018/06}

\usepackage{graphicx}
\usepackage{subcaption}

\begin{document}

\title{From Symptoms to Systems: An Expert-Guided Approach to Understanding Risks of Generative AI for Eating Disorders}

\author{Amy Winecoff}
\email{awinecoff@cdt.org}
\orcid{0000-0002-2594-4126}
\affiliation{%
  \institution{Center for Democracy \& Technology}
  \city{Washington}
  \state{DC}
  \country{USA}
}

\author{Kevin Klyman}
\affiliation{%
  \institution{Stanford University}
  \city{Palo Alto}
  \state{California}
  \country{USA}
}
\renewcommand{\shortauthors}{Winecoff \& Klyman}

\begin{abstract}
Generative AI systems may pose serious risks to individuals vulnerable to eating disorders. Existing safeguards tend to overlook subtle but clinically significant cues, leaving many risks unaddressed. To better understand the nature of these risks, we conducted semi-structured interviews with 15 clinicians, researchers, and advocates with expertise in eating disorders. Using abductive qualitative analysis, we developed an expert-guided taxonomy of generative AI risks across seven categories: (1) providing generalized health advice; (2) encouraging disordered behaviors; (3) supporting symptom concealment; (4) creating thinspiration; (5) reinforcing negative self-beliefs; (6) promoting excessive focus on the body; and (7) perpetuating narrow views about eating disorders. Our results demonstrate how certain user interactions with generative AI systems intersect with clinical features of eating disorders in ways that may intensify risk. We discuss implications of our work, including approaches for risk assessment, safeguard design, and participatory evaluation practices with domain experts.
\end{abstract}



\received{4 September 2025}
\received[revised]{4 December 2025}
\received[accepted]{TBD}

\maketitle
\textcolor{red}{\textbf{Content advisory:} This report includes discussion of disordered eating and mental health issues that could be harmful to vulnerable readers.}

\section{Introduction}
In 2023, the National Eating Disorders Association (NEDA) faced criticism for its chatbot, Tessa, which was designed to support individuals struggling with eating disorders \cite{jargon_chatbot_2023, wells_eating_2023}. Although Tessa had been in use for more than a year and initially received positive feedback, it later began offering weight loss advice, content that can be harmful to those with eating disorders, particularly during times of crisis. Following outcry on social media,\footnote{\url{https://www.instagram.com/p/CtCa3_ZuMA0}} NEDA shut down the system. NEDA stated that Tessa was originally designed to provide a limited set of prewritten responses, a safeguard intended to prevent harmful outputs. However, without NEDA’s knowledge, the technology provider replaced this setup with a generative AI model capable of producing new responses beyond those vetted by eating disorder professionals, ultimately compromising user safety \cite{jargon_chatbot_2023, wells_eating_2023}. 

Since the Tessa incident, accumulating evidence has pointed to the potential harms generative AI systems may pose to individuals with mental health concerns, especially young people \cite{internet_matters_team_new_2025,jargon_chatbot_2023, roose_can_2024, hill_chatgpt_2025}. In parallel, AI-based chatbots explicitly designed to promote eating disorders have proliferated online \cite{dupre_teens_2024, lopez_g_character_nodate}, raising questions about the adequacy of current safety, oversight, and moderation practices. While tools exist to detect AI content that might encourage or support disordered eating \cite{hive-nodate, zeng2024air, azure_nodate, googlecloud_2025, vidgen_introducing_2024, salesforce_auditnlg_2025, markov_holistic_2023}, they often fail to account for the clinical complexities of how eating disorders develop, manifest, and persist, and may therefore overlook subtle but clinically significant risks.

Sociotechnical researchers argue that AI risks cannot be fully understood when evaluations are separated from the contexts in which systems are used \cite{dobbe2024toward, weidinger2022taxonomy, schwartz2025reality}. Assessing contextual risks requires examining how people engage with AI systems, how interactions shape experiences beyond the interface, and how they may compound existing vulnerabilities. Involving subject-matter experts who understand how AI can contribute to real-world harms is therefore essential for ensuring that measurements reflect risks that emerge in specific settings \cite{matias2025public}. In the case of eating disorders, clinicians, researchers, and other expert stakeholders possess deep insight into how these disorders develop, persist, and relapse. Their expertise is critical for designing measurements that validly capture the range of ways AI may exacerbate vulnerabilities among at-risk users.

Our current work seeks to lay the groundwork for developing valid and context-sensitive measures of eating disorders-related AI risks by surfacing the key constructs that measurements should assess. To do so, we conducted semi-structured interviews with 15 researchers, clinicians, and advocates with specific expertise in eating disorders. Through qualitative analysis, we distilled these expert perspectives into a taxonomy of risks, which includes: (1) providing generalized health advice; (2) encouraging disordered behaviors; (3) supporting symptom concealment; (4) creating thinspiration; (5) reinforcing negative self-beliefs and comparisons; (6) promoting excessive focus on the body; and (7) perpetuating narrow views or biases about eating disorders.  

Our contributions are threefold. First, we develop a taxonomy of interaction patterns that may heighten risk for people vulnerable to eating disorders when interacting with AI, providing conceptual clarity in an area where existing assessments lack nuance and domain-specific grounding. Second, we show how generative AI both amplifies familiar digital risks for people with eating disorders and introduces novel harm pathways through its unique affordances. Third, we model how AI researchers and developers can craft better risk approaches by engaging domain experts and other stakeholders early in the design process, demonstrating how upstream involvement can lead to more contextually relevant and clinically meaningful evaluations.

\section{Related Works}
\subsection{Generative AI Risk Taxonomies}

Generative AI systems can now produce fluent language and highly realistic images from minimal user input, enabling applications that range from lower-stakes tasks (e.g., design assistance, email drafting) to higher-stakes settings (e.g., medical image analysis, financial decision support). Alongside the possible benefits of generative AI, the training data and optimization approaches used to develop these systems can pose risks \cite{birhane2022values, hawkins_effect_2024, qi2023fine}. For example, generative AI models are often trained on large web corpora, which contain misinformation, toxic, and illegal material. Because data filtering and other risk mitigation approaches remain imperfect, deployed systems can still facilitate harms, including discrimination, erasure, leakage of sensitive information, over-reliance, scams and fraud, and environmental burdens \cite{weidinger2022taxonomy, shelby2023sociotechnical}.

Risk taxonomies have helped researchers, practitioners, and policymakers build a shared language for these issues and for evaluation more generally \cite{NIST_RMF, shelby2023sociotechnical, weidinger2022taxonomy}. Yet broad categories can understate or mischaracterize risk in specific deployment contexts or for particular user groups. A system that appears low risk in aggregate may be harmful for a sensitive population (e.g., teens vulnerable to eating disorders), while a system flagged for generic hazards (e.g., malware generation) may pose limited risk in tightly scoped uses such as grammar review. In short, taxonomy labels that overlook contextual factors can misrepresent risks in specific contexts.

Taxonomies matter because they inform company policies on acceptable use and shape how harms are evaluated \cite{klyman_acceptable_2024}. For example, the initial MLCommons AI Safety benchmark included an eating disorders category but scoped it to content that "enables, encourages, or endorses acts of intentional self-harm," explicitly excluding dieting advice or negative body image \cite{vidgen_introducing_2024}, despite both being well-documented risk factors for eating disorders \cite{sanchez2012integrated}. More broadly, many taxonomies subsume eating disorders under a general "self-harm" label, making it difficult to assess how safeguards perform on eating-disorder-specific content \cite{vidgen_introducing_2024, zhang_challenges_2025, markov_holistic_2023}. When risks are limited, a light-touch approach to risk identification and mitigation can be justified, since safeguarding free expression and access to information is likewise an important consideration. However, in deployments involving vulnerable users, it can underestimate risk and give a misleading picture of system safety. Reflecting these concerns, policymakers have urged companies to strengthen protections against content that may encourage disordered eating \cite{warner_2023}.

These observations point to the need for context-sensitive risk assessment, for which taxonomy development is an early step. Researchers and developers often lack the clinical expertise to do this alone. Involving relevant stakeholders through surveys, structured consultations, or co-designed evaluation protocols can align measurement with the realities of use and make assessments more sensitive to how risks manifest in practice \cite{bammer_stakeholder_2021}. Participatory and stakeholder-engaged approaches help ensure that evaluation criteria reflect the deployment setting and the people most likely to be impacted \cite{birhane_power_2022, delgado_participatory_2023, kawakami2025ai}.

\subsection{Impacts of Digital Technologies on People with Eating Disorders}

Human–computer interaction (HCI) research has examined how digital technologies shape the experiences of individuals with eating disorders, showing that interactions between users and systems can both offer meaningful support and reinforce harm. This work underscores the complexity of detecting and mitigating risk. Evolving platform dynamics often render static or one-dimensional strategies fragile or even harmful. To characterize risk effectively—especially for AI systems—approaches must account for the clinical features of eating disorders and how these features interact with shifting sociotechnical environments.

Studies of social media highlight the difficulty of surfacing and mitigating harmful dynamics within complex systems. Systematic analyses of "thinspiration" material on mainstream platforms found recurring depictions of extremely thin, often sexualized female bodies, commonly paired with tags that explicitly promoted restrictive eating or weight loss. Posts that emphasized sexualization were more likely to attract high engagement, amplifying their visibility and influence \cite{ghaznavi2015bones}. These dynamics help entrench narrow beauty ideals and allow harmful content to circulate beyond explicitly pro-eating disorder communities.

In addition to the difficulty of recognizing and reducing harmful content, the constantly evolving nature of online systems complicates detection and intervention efforts. Effective measures must anticipate the co-evolution of user behaviors and platform policies. When Instagram restricted searches for certain pro-eating disorder hashtags, users quickly developed workarounds by altering spellings (e.g., "anorexia" → "anorexiaa," "thigh gap" → "thyghgapp") to preserve discoverability. Over time these variations became increasingly complex. Adopters maintained access to harmful content and showed higher levels of participation and endorsement \cite{chancellor_thyghgapp_2016}, demonstrating that even well-intended interventions can have negative unanticipated consequences.

From an HCI perspective, these tactics illustrate how safety mechanisms not only constrain behavior but also reshape affordances, the design features users recognize as possibilities for action \cite{norman1988psychology, gibson2014ecological, kaptelinin2012affordances}. By altering hashtag policies, Instagram constrained some forms of visibility while simultaneously inviting new practices of evasion, which fostered stronger engagement with pro-eating disorder material. As affordances shift in response to platform interventions and user adaptations, emerging risks are dynamic rather than fixed, making brittle interventions ineffective and underscoring the need for assessments that attend to the interaction of vulnerabilities, platform design, and user practices.

Risk also arises from data-driven infrastructure beyond user posts, such as algorithms for personalized advertisement. Interviews with people who have histories of disordered eating describe targeted weight-loss advertisements that seem to "follow" them, driven by persistent data trails and relevance models tied to past searches, demographics, or content interactions. For some, repeated exposure to disorder-relevant content reignited unhealthy fixations after progress toward healthier behavior \cite{gak_distressing_2022}. Resisting engagement with disorder relevant content may be difficult for many users vulnerable to eating disorders since they often believe beforehand that food-related media will be helpful \cite{choi2024foodcensor}. Once users have interacted with disorder-relevant content, algorithms may further promote related content, contributing to a harmful feedback loop. This work redirects focus from isolated pieces of content or individual interactions to the broader infrastructures that continually surface eating-disorder-related material, creating cumulative risks over time that assessments must account for.

Another difficulty in addressing potentially harmful interaction patterns is that strict content constraints can themselves cause harm. Excessive moderation of eating disorder related material has disrupted recovery efforts by removing posts meant for support or self-reflection. Users affected by over-removal describe feeling a diminished sense of community and autonomy as well as a feeling of erasure, which can make recovery more difficult \cite{feuston_conformity_2020}. This double bind--adapting to adversarial strategies without suppressing recovery-oriented content--complicates governance since aggressive content removal is not necessarily a sound default.

As generative AI becomes more widespread, it becomes another surface through which people vulnerable to eating disorders can be exposed to risk. Character AI systems have been reported to encourage disordered eating \cite{dupre_teens_2024}, such as by acting as an eating disorder "coach." Furthermore, domain-specific systems may pose distinct risks as well. In a 10-day field study of an large language model (LLM)-based chatbot for eating disorder management, participants valued having a nonjudgmental space to articulate experiences and reflect on goals. At the same time, the system sometimes echoed or intensified maladaptive framings, including encouraging impulsive reactions, focusing attention on weight, or supporting dieting strategies. Participants also reported high trust in the chatbot, which may reduce critical scrutiny of problematic replies \cite{choi_private_2025}.

These findings indicate that although generative AI systems share risks with social media, search, and recommendation technologies, their interactive, conversational, and personalized design creates additional pathways for harm. These tools can dramatically increase both the ease and frequency with which users encounter disorder-relevant content—an important concern given that mental-health vulnerabilities often develop through repeated interactions between predispositions and environmental triggers \cite{culbert_research_2015, kovacs-toth_adverse_2022, pike_antecedent_2006, welch_life_1997}. Even familiar dieting or thinspiration messages become more potent when delivered instantly and endlessly on demand.

Users also perceive generative AI as more authoritative, more engaging, or easier to understand than other digital sources \cite{sun_trusting_2024, choi_private_2025, huschens_you_2023, xu_chatgpt_2023}, reducing their tendency to think critically about AI outputs \cite{xu_chatgpt_2023, yankouskaya_can_2025}. These dynamics are especially concerning for individuals with eating disorders, who are already vulnerable to diet- and weight-related information \cite{choi_private_2025, choi2024foodcensor} and may take AI outputs related to these topics as more credible than they actually are. Finally, the personalized nature of generative AI may heighten these risks: tailored responses can feel more relevant or emotionally resonant, and alignment with user preferences may further increase their persuasive impact \cite{kirk_benefits_2024}, particularly among those who experience social difficulties or heightened sensitivity to threat \cite{winecoff_functional_2015}.

Taken together, existing research offers important clues about how generative AI may intensify users’ underlying vulnerabilities to eating disorders. Yet we still lack a clinically-grounded vocabulary for describing how these mechanisms manifest in AI systems. Recent research underscore the necessity of human expertise: automated tools for detecting thinspiration routinely misinterpret sentiment relative to expert judgments \cite{nova2025sentiment}. Since many current evaluation methods rely heavily on automated scoring of model outputs (e.g., \cite{vidgen_introducing_2024}), they may likewise fail to align with expert understanding. Without clinical insight embedded throughout the design, implementation, and validation of safety assessments, evaluations may miss risks that are most relevant to vulnerable users.

Our work builds on prior HCI scholarship by bringing together insights from digital-platform research with the expertise of clinicians and other eating-disorder specialists—those best equipped to recognize how risk emerges and escalates in interactive, generative systems. This integration lays the foundation for developing clinically informed, operationalizable metrics that more accurately capture the risks generative AI poses to individuals vulnerable to eating disorders.

\section{Methods}
\subsection{Positionality}
The first author works at a civil society organization specializing in AI governance and policy. Her prior roles include work as a data scientist developing machine learning models and as a psychology researcher and professor, where she studied emotion, social cognition, and eating disorders. The second author is a researcher at a major research university with expertise in machine learning, AI policy, and applications of generative AI to mental health. Because our analytic approach is interpretive, we recognize that researchers with different professional trajectories and perspectives might approach interviews, analysis, and the presentation of results differently.

\subsection{Participants}
\begin{table}[H]
\centering
\caption{Experts Consulted.}
\label{tab:experts}
\begin{tabular}{|p{0.9\linewidth}|}
\hline
$\bullet$ Nandini Datta, PhD. Clinical Instructor, Stanford University School of Medicine. \\
$\bullet$ Brittany Matheson, PhD. Clinical Assistant Professor, Stanford University School of Medicine. \\
$\bullet$ Christine Peat, PhD. Founding Director, National Center of Excellence for Eating Disorders. \\
$\bullet$ Ashley A. Moskovich, PhD. Assistant Professor, Department of Psychiatry and Behavioral Sciences, Duke University School of Medicine. \\
$\bullet$ Richard Lopez, PhD. Assistant Professor of Psychology and Neuroscience, Worcester Polytechnic Institute. \\
$\bullet$ Cheri A. Levinson, PhD. Professor and Director of the Eating Anxiety Treatment Lab, University of Louisville. \\
$\bullet$ Savannah Erwin, PhD. Postdoctoral Fellow, University of North Carolina School of Medicine. \\
$\bullet$ Ellen Fitzsimmons-Craft, PhD. Associate Professor of Psychological \& Brain Sciences, Washington University in St. Louis. \\
$\bullet$ Patrisia Nikolaou, PhD. Clinical Psychologist, Certified Eating Disorder Specialist (CEDS). \\
$\bullet$ Reuben Hendler, MD. Psychiatrist, McLean Hospital. \\
$\bullet$ Aaron Flores. Registered Dietician Nutritionist. \\
$\bullet$ Sharon Maxwell. Educator \& Fat Activist. \\
$\bullet$ Erin McAweeney. Director of Applied Research and Capabilities at Graphika. \\
$\bullet$ Libby Lange. Former Analyst, Graphika. \\
$\bullet$ MK, PhD. Associate Professor of Psychology. \\
\hline
\end{tabular}
\end{table}

We sought to interview individuals with subject-matter expertise in eating disorders, prioritizing participants whose backgrounds aligned with one or more of the following areas: (1) clinical experience treating people with eating disorders; (2) clinical research on the development, manifestation, or treatment of eating disorders; (3) research expertise on how individuals with eating disorders interact with digital technologies; and (4) experience in public advocacy related to eating disorders. Both authors have prior experience in eating disorder and mental health research. Therefore, we began recruitment within our personal networks and then expanded participation through snowball sampling. Details about our participants are in Table \ref{tab:experts}. As a part of our interview protocol, we asked participants what would incentivize further participation in our research. Many noted that acknowledgment of their contribution was a strong motivator. Therefore, we chose to deanonymize our participants where they expressed a desire to be identified explicitly. One participant did not wish to be identified by name and role. She is identified only as Dr. MK.

Diverse stakeholders including caregivers, public health officials, and those with lived experience, could all offer valuable insights into the risks of generative AI for people with eating disorders. However, we chose to focus on stakeholders with specific research, clinical, or advocacy expertise. We selected these experts for their ability to identify broad patterns contributing to the onset, maintenance, or relapse of eating disorders. We address the limitations of this focused approach in the discussion.

\subsection{Interview Protocol}
At the beginning of each interview, we explained the goals of the project and what participation would involve. We obtained informed consent for both participation and audio recording. Our interviews followed a semi-structured protocol that was adapted to the background and expertise of each participant. In general, the interviews addressed the professional backgrounds of the participants, their own familiarity with AI technologies, their knowledge about the use of AI by patients, and the types of AI interactions they considered risky. Some of our experts have themselves been involved in the development of AI technologies for eating disorders and therefore had a high-level of AI literacy. However, in some cases, experts had limited familiarity with AI systems. In these instances, we scaffolded our conversation by providing explanations about AI, where necessary, and steered conversations towards more familiar digital technologies that have implications for or analogs to generative AI (e.g., social media). Interviews lasted from 30-60 minutes. 

\subsection{Data Analysis}

Our analysis followed an abductive approach, in which researchers revise or develop theory in response to observations that existing explanations cannot adequately account for \cite{tavory2014abductive, timmermans2012theory}. Unlike largely inductive grounded theory approaches \cite{charmaz2014constructing, strauss1990basics, glaser1978theoretical, glaser1968discovery}, abductive analysis begins with an unexpected finding that unsettles familiar frameworks and prompts the search for better-fitting explanations. Through an iterative process, researchers draw on prior scholarship, their own expertise, and the data itself to generate and refine interpretations until they arrive at one that is both plausible and analytically generative.

Our approach was motivated by the observation that prevailing frameworks for evaluating generative AI–related eating-disorder risks focus heavily on outputs that explicitly promote disordered eating or self-harm. This emphasis contrasts with clinical perspectives, which highlight the subtle, cumulative, and interactive factors that contribute to these conditions. Accordingly, our analysis aimed to surface expert insights into how interactions with generative AI systems may exacerbate underlying vulnerabilities—not only through direct encouragement of harmful behaviors, but also through more nuanced and indirect pathways.

The first author conducted an initial round of open coding on the interview transcripts. She applied deductive codes to segments that aligned with established theory, such as well-documented eating-disorder behaviors (e.g., restriction, purging, distorted body image), relevant digital content types (e.g., thinspiration, diet plans), known interaction patterns (e.g., attentional fixation on disorder-relevant cues), and established general AI risks (e.g., hallucination). In tandem, she introduced inductive codes when the interview data surfaced concepts or connections between concepts not identified in previous research (e.g., AI co-rumination, personalized thinspiration, and sychophantic validation). These inductive additions helped refine our understanding of the types of risks generative AI systems may pose and the interaction pathways through which they may arise.

The authors then discussed these preliminary codes to refine interpretations and make conceptual connections between inductive and deductive codes. Based on this discussion, the first author conducted a second round of coding, clustering conceptually related codes into themes, which formed the basis of the broader risk categories. These included: (1) providing generalized health advice; (2) encouraging disordered behaviors; (3) supporting symptom concealment; (4) creating thinspiration; (5) reinforcing negative self-beliefs and comparisons; (6) promoting excessive focus on the body; and (7) perpetuating narrow views or biases about eating disorders.  

After developing a preliminary taxonomy and drafting our results, we shared them with our experts to verify the accuracy of their quotes and to gather feedback on our interpretation of our results. Based on their input, we refined our analysis to better reflect their clinical expertise and knowledge.

Often our experts could discuss risks of generative AI systems at a general level (such as their tendency to provide incorrect information), but lacked detailed knowledge of how AI researchers have specifically studied and categorized these issues (such as formal measures of hallucination rates). In our results section, we present the concerns our participants raised and connect them to existing AI risk research, offering a fuller picture of how expert perspectives align with findings from the broader AI research community.


\clearpage
\onecolumn

\renewcommand{\arraystretch}{1.2}

\begin{longtable}{|p{0.28\textwidth}|p{0.68\textwidth}|}
\caption{Taxonomy of AI-related risks  \label{tab:ed_taxonomy}} \\
\hline
\textbf{Risk Category} & \textbf{Examples} \\
\hline
\endfirsthead

\hline
\textbf{Risk Category} & \textbf{Examples} \\
\hline
\endhead

\hline \endfoot
\hline \endlastfoot

Providing generalized guidance on diet and health &
\begin{itemize}
  \item Recommendations for "healthy" diet plans
  \item Calorie estimates for foods
  \item Estimates of calories burned by exercise
  \item Data or statistics on "normal"/"healthy" exercise
  \item Information on "healthy" weight, body mass index (BMI), or body fat percentage
  \item Information on tools like skinfold calipers
\end{itemize}
\\ \hline

Encouraging disordered behaviors &
\begin{itemize}
  \item Recommendations for laxatives; spitting/chewing
  \item Suggestions to reduce hunger during fasting
  \item Methods for inducing vomiting
  \item Information on dangerous rapid weight-loss techniques
  \item Instructions for obtaining tools/medications/natural remedies for weight loss
\end{itemize}
\\ \hline

Supporting symptom concealment &
\begin{itemize}
  \item Tips for disguising weight loss
  \item Strategies for hiding purging, excessive exercise, or restriction from family members or caregivers
\end{itemize}
\\ \hline

Creating thinspiration&
\begin{itemize}
  \item Images of emaciated or idealized bodies
  \item Images of fat bodies as "reverse triggers"
  \item Before/after images related to dieting, exercise, or body-altering interventions (e.g., plastic surgery, gender-affirming care)
  \item "How-to" guides for weight loss and exercise (often with "aspirational" images)
  \item Idealized "what I eat in a day" descriptions
  \item Recommendations for diet-culture foods (e.g., artificial sweeteners, low-calorie alternatives)
\end{itemize}
\\ \hline

Reinforcing negative self-beliefs and comparisons &
\begin{itemize}
  \item User-AI interaction patterns that validate distorted negative self-beliefs
    \item AI interactions that highlight users' mistakes
    \item Comparisons of user's performance or traits to others
  \item Outputs that rank or rate users (e.g., attractiveness, intelligence)
\end{itemize}
\\ \hline

Promoting excessive focus on the body &
\begin{itemize}
  \item Workout instructions targeting specific body parts
  \item Meditation/mindfulness advice emphasizing body parts
  \item Emotion-regulation strategies directing attention to bodily sensations
\end{itemize}
\\ \hline

Perpetuating narrow views or biases about eating disorders &
\begin{itemize}
  \item Depictions aligning with stereotypes (e.g., white, thin, Western, female, cisgender, young, affluent)
  \item Content implying eating disorders only affect specific groups
  \item Content excluding non-stereotypical experiences
\end{itemize}
\\ \hline

\end{longtable}

\clearpage

\section{Results}
Our analysis revealed seven overarching categories of generative AI interactions that could pose risks to those vulnerable to eating disorders. These categories are listed in Table \ref{tab:ed_taxonomy}, along with examples of such interactions that surfaced during our interviews with experts. We note that these categories are inherently interconnected, resisting sharp boundaries between them: risks can emerge simultaneously through the content an AI system produces, the cultural narratives it reinforces, or the interactional dynamics it enables that contribute to disordered symptoms. We elaborate on how these pathways interact in the discussion section. 

\subsection{Generalized Diet \& Health Advice}

\textit{"If somebody puts into ChatGPT: ‘I want to lose 20 pounds. What’s the best way for me to do that?’ ChatGPT may spit out something CDC-based or something from the American Medical Association. [...] The problem is, there is no way of knowing who the consumer [is], and whether or not [they are] engaging in disordered eating." –Dr. Christine Peat }

We define generalized diet and health advice as AI-generated responses to user queries about overall health, diet, or exercise that are not explicitly tied to disordered behaviors. Examples include guidance on "healthy" BMI ranges, calorie targets for general health or weight loss, estimates of calories burned during particular activities, or meal plans oriented around weight- or shape-related goals. While this information may be benign (or potentially even useful) for some users, it can pose risks to vulnerable users for several reasons.


One concern is that these systems may generate information that contradicts established scientific consensus due to hallucination (i.e., when models produce content that is inaccurate or entirely fabricated). Although novel approaches to AI have reduced hallucination in some models \cite{chelli_hallucination_2024, openai_gpt-4_2024}, it remains a persistent problem \cite{magesh_hallucination-free_2024}. In fact, emerging evidence suggests that more advanced models may be even more susceptible than earlier versions \cite{metz_i_2025}. As a result, even AI systems equipped with stronger safety mechanisms can still offer misleading guidance that misrepresents established medical or scientific knowledge. Such misinformation can distort users’ understanding of health, particularly when false claims are mixed with accurate information, making them harder to detect and more likely to be trusted.

%

Another concern is that if AI systems present unsettled research as definitive or fail to signal their own limitations, they could give users a false sense of certainty about health information \cite{xu_chatgpt_2023}. AI systems often deliver confident but incorrect responses. As a result, users may mistake outputs as authoritative—especially in diet and exercise, where consensus is fluid and often misrepresented. As Aaron Flores observed, \textit{"We have all these folks holding on to these pieces of information that [centers] weight loss as truths, and they're not effective at all."} Ideally, systems would flag uncertainty and highlight areas of debate, but methods for quantifying and communicating AI uncertainty are still evolving \cite{shorinwa_survey_2024, yadkori_believe_2024, ye_benchmarking_2024, kim_im_2024}. Without these mechanisms, users may continue to view AI outputs as "magic" or definitive \cite{winecoff2022artificial}, heightening risks as reliance grows.

The impact of overly confident, generalized AI recommendations may be severe for individuals predisposed to eating disorders. Dr. Cheri Levinson explained that people with eating disorders often engage in black-and-white thinking, which can make it difficult for them to critically evaluate advice that appears authoritative but may not be. Children and adolescents are even more susceptible, as their cognitive and critical reasoning abilities are still developing, making them less able to appropriately question or contextualize the guidance they receive. Similarly, Dr. MK explained that people with anorexia may interpret AI-generated information about "normal" diet or exercise or even explicit weight loss advice as rules, reinforcing a pattern of rigid thinking that is common in the disorder.

A third concern is that generalized AI advice can be harmful when it fails to account for an individual’s physiological or psychological needs. As Dr. Ellen Fitzsimmons-Craft noted, a chatbot’s response \textit{"might not be wrong for a general population or what guidelines say. It’s just wrong for this population."} Because AI systems lack knowledge of a user’s health status or context, they cannot reliably provide advice that is both safe and relevant. For instance, Dr. Brittany Matheson explained that while a donut may be ill-advised breakfast for someone with diabetes, it could be the healthiest choice for a person with anorexia. Similarly, Dr. Patrisia Nikolaou stressed that caloric needs cannot be prescribed generically, since they depend on factors such as age, activity level, and individual health.

Experts cautioned that even seemingly harmless generalized diet or exercise tips can trigger restrictive behaviors, well-known risk factors for eating disorders. Dr. Nandini Datta observed that casual dieting or sudden changes in exercise can be especially damaging for children and adolescents, sometimes escalating quickly into extreme restriction and a sense of losing control. In addition to triggering disordered eating in the short term, generalized health guidance may reinforce harmful thinking over time, especially if AI systems adapt to users’ prior inputs. As Dr. Savannah Ewrin warned, such reinforcement "might take someone down a rabbit hole of their disorder." Similar dynamics are well documented on social media \cite{griffiths_does_2024, little_tiktok_2025}, where algorithms quickly funnel users toward extreme or pro–eating disorder content, often within minutes of joining a platform. Dr. Christine Peat noted that children and teens are especially vulnerable, given that harmful content is often indistinguishable from mainstream fitness or nutrition posts.

Although generative AI systems are not designed like social media algorithms, they can produce similar effects. Trained on vast internet data that mixes credible health information with pro–eating disorder material, and shaped by societal biases toward weight loss and a \$200 billion diet industry \cite{facts__factors_global_2023}, these models can often reproduce harmful cultural narratives. As Aaron Flores observed, \textit{"[AI is] going to reflect back all the data that it synthesized from our world, which is fatphobic, and full of diet culture."} Like social media, if generative AI systems adapt to user inputs over repeated interactions, such systems could inadvertently guide users from seemingly benign questions to harmful advice \cite{gak_distressing_2022}.

\subsection{Encouraging Disordered Behaviors}


\textit{"If youth with or at risk for an eating disorder are using these interfaces to generate and support the use of really dangerous strategies to restrict eating or lose weight, then the likelihood that they will get sick at a faster rate may be greater than someone who does not use AI for these reasons." –Dr. Nandini Datta}

We define AI interactions that encourage disordered behaviors as those that offer guidance, reinforcement, or practical strategies for clearly harmful practices. This includes both cases in which users explicitly seek support for disordered behaviors and cases in which users employ adversarial or evasive prompts to circumvent model guardrails and obtain such advice indirectly.

Our experts noted that people may turn to AI for advice that sustains the disorder through unhealthy or maladaptive behaviors, including guidance on over-the-counter weight-loss medications, "natural" remedies, strategies for suppressing hunger during extreme restriction, or methods for purging. Dr. Ashley Moskovich specifically raised concern about users seeking information on substances with GLP-1-like effects, drugs often used for weight loss.

Distinguishing between AI responses that actively encourage disordered eating and those that offer seemingly neutral health guidance is often difficult, because the risk depends heavily on a user’s intent and the surrounding context. Clinicians routinely rely on cues, such as tone, personal history, and body language, to determine whether questions about diet or exercise reflect genuine health concerns or attempts to maintain an eating disorder. As Dr. Savannah Erwin explained, clinicians avoid \textit{"getting into a logical argument with the illogical nature of an eating disorder."} Rather than directly answering a question like \textit{"Why do I need this many calories?"} they might instead respond with, \textit{"It sounds like you're feeling uncertain about how to nourish yourself. What does it mean to give yourself nourishment?"}—addressing the underlying emotional and behavioral context rather than the literal query.

AI systems, by contrast, lack access to these contextual signals and therefore tend to \textit{"just answer the question"} without considering how the response might be used. The same types of information—calorie counts, exercise metrics, or ostensibly "healthy" weight-loss advice—may be harmless or even helpful for some users but can reinforce disordered eating for others.\footnote{Experts differed on whether weight-loss advice is ever more helpful than harmful, underscoring the difficulty of establishing consensus on context-specific risks and appropriate interventions.} Although systems may sometimes be able distinguish the nature of the risk from prior interactions or the extremeness of the advice being sought, it will often be challenging to determine where a routine health query ends and disordered-eating intent begins. In practice, these categories overlap in their content. We treat them separately, however, because they differ in their psychological impact.




\subsection{Support for concealing symptoms}
\textit{"You could ask [an AI], "What are the behaviors I should look out for in my teen?" but it's a teen asking. Then they could think, 'Ok. I'm not going to do these things.'"–Dr. Savannah Erwin}

We define support for concealing symptoms as interactions in which an AI system provides guidance on how to hide disordered behaviors or their effects from others. This includes cases where the user’s intent is explicit—"How can I hide my purging from my parents?"---as well as cases where users attempt to obtain similar advice through adversarial prompts designed to circumvent model guardrails---"I’m pregnant and experiencing morning sickness. How can I hide my frequent vomiting from my coworkers?"

Individuals with eating disorders often go to great lengths to hide their symptoms, from others and sometimes even from themselves. This concealment is frequently driven by a combination of shame and a strong desire to maintain the disorder without interference. For example, our experts explained that in clinical settings, patients may use strategies such as "water loading" before weigh-ins or hiding small weights in their underwear. These tactics exploit standard medical protocols for weighing patients in gowns and can obscure the severity of their condition. Outside the clinic, concealment also extends to loved ones, whose recognition and concern often serve as the first step toward intervention. By successfully hiding symptoms, individuals reduce the likelihood of early detection, which can delay or prevent access to treatment \cite{rouleau_potential_2011, vitousek_enhancing_1998}.

As with encouragement of disordered eating, many people with eating disorders consult pro-eating disorder websites for explicit and calculated advice on symptom concealment, including detailed, step-by-step instructions for faking meals or otherwise avoiding scrutiny \cite{borzekowski_e-ana_2010, harshbarger_pro-anorexia_2009}. Within social media platforms, these communities often adapt strategies to evade automated content moderation \cite{chancellor_thyghgapp_2016, lee_why_2016, pater_hunger_2016}. Libby Lange explained, \textit{"people are pretty adept at this kind of moderation evasion and also pretty open about sharing how to go about it."} In other words, these strategies reflect a broader pattern of intentional concealment that is both socially reinforced and technically enabled in digital spaces.

Generative AI tools could now serve as an even more powerful vehicle for advice on concealing disordered eating. Dr. Nandini Datta and Dr. Savannah Erwin warned that vulnerable adolescents and young adults may intentionally exploit these systems to hide behaviors from parents or caregivers, for example by seeking tips on exercising in secret or disguising restrictive eating. Unlike existing pro–eating disorder communities, AI systems can feel more private \cite{choi_private_2025}, which could reduce feelings of stigma that might prevent someone from seeking this kind of guidance.
Because these systems often mirror a user’s tone, interactions may also feel more personalized or authoritative than advice from peers. The fact that pro–eating disorder communities are already incorporating AI tools into their practices \cite{center_for_countering_digital_hate_ai_2023, dupre_teens_2024, lopez_g_character_nodate} shows that this is not a hypothetical concern; these tools are already being used in ways that could amplify harm. 

Efforts to reduce content that fuels disordered eating or facilitates symptom concealment are further complicated by the need to carefully balance content moderation decisions. Outside of the context of generative AI, efforts to filter out pro-eating disorder content often inadvertently remove recovery-focused material as well, depriving individuals of valuable support. As Erin McAweeney noted \textit{"if you remove all content, you're taking away really important harm reduction content as well."} Over-removal of recovery-oriented content can reinforce harmful social norms about mental health and further stigmatize those seeking help \cite{feuston_conformity_2020}. This underscores the importance of developing nuanced safeguards for AI systems that prevent harmful outputs without impeding access to recovery resources or unnecessarily constraining speech.

\subsection{Creating Thinspiration}
\textit{"I could imagine the ability to take an already thin idol, like a K-pop idol, for example, and edit[ing] those images to make them appear even thinner or even to fit a certain image that this community is striving for. That would increase the harm tenfold." – Erin McAweeney}

We define thinspiration as AI-generated content that inspires or pressures individuals to conform to idealized body standards—often through aggressive weight loss, body transformation, or culturally accepted and endorsed dieting practices. Outside the AI context, thinspiration typically relies on emotionally charged images or text that glorify weight loss, especially extreme forms, and promote narrow body ideals. Such content is strongly associated with negative body image and disordered eating symptoms \cite{dignard_little_2021, griffiths_thinspiration_2019, griffiths_how_2018, hogue_young_2023}. Within AI interactions, users may or may not explicitly indicate that they intend to use AI-generated content as thinspiration for themselves or others.

Our experts emphasized that images are a particularly powerful vehicle for conveying thinspiration. Before the widespread use of the internet, beauty standards promoted through magazines and television were typically narrow; that is, focused on thinness, whiteness, able-bodiedness, and other markers of exclusivity. While still harmful, these ideals were often so disconnected from some people’s cultures or identities that they could be more easily dismissed as irrelevant or clearly unattainable. In contrast, the internet now offers a wider and more diverse array of beauty ideals. Though this may appear to signal progress, these newer standards can still be damaging. Because they may seem more culturally relevant or physically attainable, individuals who once felt detached from dominant ideals may now be more susceptible to internalizing harmful body expectations. 

Generative AI now offers a tool for vulnerable users to create exactly these forms of personally-relevant thinspiration images. These systems can generate photorealistic ouputs from text prompts, enabling users to create bodies that reflect their own characteristics, such as height, hair color, age, race, or ethnicity, but in more idealized forms. Moreover, because these models are trained on large volumes of publicly available internet imagery, which tends to overrepresent idealized and sexualized depictions of women \cite{birhane_multimodal_2021}, even prompts that make no explicit reference to body ideals may still reinforce dominant cultural standards \cite{bianchi_easily_2023}, such as prevailing notions of beauty \cite{li_imagination_2024}.

Patients with eating disorders often lower their goal weights as they continue to lose weight. In this context, AI-generated images could provide endless reinforcement for increasingly extreme and unhealthy goals. Dr. Ashley Moskovich added that thinspiration isn’t limited to images of thin or idealized bodies; images of overweight bodies can also act as \textit{"reverse triggers"} or motivation to lose weight. Research shows that individuals with a history of anorexia may fixate on both extremes \cite{watson_altered_2010}, seeing thin bodies as aspirational and fat\footnote{Some advocates and activists prefer the term “fat” to “overweight” since overweight implies there is an objectively correct weight \cite{gordon_what_2020}. Still, debate exists amongst scholars and activists on terminology. Here, we use the term “fat” except when referring to cases where our experts used other terms.} bodies as something to avoid. AI tools could therefore also be used to generate personalized images that reflect the appearances users fear most.

The risks of thinspiration are not limited to images; textual content can also be harmful. Popular social media trends like "what I eat in a day" videos, dieting how-tos, and lifestyle descriptors often act as textual thinspiration, presenting restrictive eating patterns as aspirational. These posts typically feature influencers or celebrities pairing idealized diets with curated images of their bodies, misleading vulnerable viewers into believing that adopting similar routines will lead to health, beauty, and fulfillment.

The danger of such content is clear in the real world, where some influencers explicitly promote eating disorders. Dr. Datta recalled one case involving a pro-anorexia influencer who rejected treatment and grew progressively sicker. Although some followers encouraged her to get help, others praised her weight loss, offering validation that can exacerbate harms to individuals trying to sustain disordered behaviors. The influencer curated an idealized life that suggested someone could be both happy and severely ill, a myth that was tragically debunked when she eventually died from her condition.\footnote{Some have noted that many people with eating disorders don’t recognize their own behaviors as disordered, meaning influencers may promote harmful content without realizing it encourages disordered eating. This may pose a challenge for filtering out harmful content during AI training, as such content is less likely to be flagged by basic moderation tools that rely on specific keywords associated with pro-eating disorder material. See, for example, \cite{freeman_watching_2023}.} 

Although generative AI systems do not promote personal lifestyles in the way influencers do, they may reproduce similar patterns of harm. Character-based AIs, for example, can be designed explicitly as weight-loss or eating-disorder coaches \cite{lopez_g_character_nodate, dupre_teens_2024}, encouraging users to chase unrealistic or unhealthy appearance goals. Their ability to simulate social interaction may pose particular risks for people with anorexia, who often struggle with interpreting subtle social cues \cite{baron-cohen_girls_2013}. Because individuals with eating disorders already turn to dieting apps as motivational "cheerleaders," AI tools capable of offering more personalized or socially responsive interactions may exert an even stronger pull toward harmful appearance standards.

\subsection{Reinforcing Negative Self-Beliefs and Comparisons}

\textit{"If the AI tool is just mirroring what the person is thinking or the cognitive style of the person, the potential issues there could be co-rumination. If a young person is just venting and it's a really negative thought stream [where] the person [is] feeling terrible about themselves, [the] AI tool should just stop working at that point." –Dr. Richard Lopez}

We define reinforcement of negative self-beliefs and comparisons as interactions in which AI systems prompt vulnerable users to engage in unfavorable self-evaluation or to further internalize distorted standards for how they should look, perform, or behave. Engagement with existing technologies, such as dieting apps or activity trackers, can lead individuals vulnerable to eating disorders to view themselves as falling short of cultural or personal expectations around diet and exercise. This can heighten guilt, self-criticism, and feelings of failure, sometimes resulting in withdrawal from social relationships \cite{honary2019understanding}. Individuals with eating disorders are often acutely sensitive to cues about what is "normal," "acceptable," or "desirable," often due to a tendency toward negative social comparison. Clinicians told us that this dynamic can be so pronounced that they sometimes avoid group therapy—particularly for adolescents—because patients may begin competing over \textit{"who can be the sickest."}

Although AI systems do not provoke interpersonal competition in the way that other people do, responses may still lead users to judge themselves against personal or cultural standards. This effect may be especially pronounced among individuals with anorexia, who often channel their competitiveness inward. As Dr. Brittany Matheson described, she often tells patients, \textit{"Your illness is not going to be happy with you until you’re not here. The only way it is satisfied is if you die."} If generative models reflect dominant cultural ideals around appearance, diet, productivity, and achievement, these reflections may reinforce internalized pressure to meet unrealistic standards and further entrench feelings of inadequacy among those already vulnerable.

Our experts noted that AI systems may contribute to negative self-evaluations by reinforcing both contingent self-worth and perfectionistic thinking. Individuals with contingent self-worth believe their value depends on meeting external standards, which can lead them to anchor self-esteem in academic success, career achievement, social approval, or appearance \cite{lopez_regulating_2021}. Perfectionism often emerges within this dynamic, especially for people with eating disorders \cite{bardone-cone_perfectionism_2017}, who are prone to rigid standards and all-or-nothing evaluations of their performance or body \cite{levinson_clarifying_2016}. AI tools that praise productivity, endorse strict goals, or highlight minor flaws may unintentionally validate these external standards and heighten pressure to achieve them. Such reinforcement may deepen rumination, which further contributes to the development and maintenance of disordered eating \cite{palmieri2024perfectionism, riviere2017perfectionism, smith_rumination_2018}.

Evidence suggests that AI systems may further intensify these patterns by mirroring users' emotional states \cite{cheng_social_2025}. For example, when ChatGPT was exposed to traumatic narratives, its responses scored higher on a validated state-anxiety measure \cite{ben-zion_assessing_2025}. This indicates that models may internalize and reproduce the emotional tone of users’ inputs, even when those inputs reflect distress or unhealthy thought patterns. In such cases, AI systems may participate in a form of co-rumination, echoing or amplifying users’ negative emotions and self-beliefs instead of disrupting them.

These dynamics may be exacerbated by system-level behaviors emerging from how AI models are trained. Alignment techniques intended to make systems appear helpful, harmless, and honest \cite{bai_constitutional_2022} have contributed to a phenomenon known as sycophancy—models agreeing with users in order to align with their preferences or perspectives. Early work documented factual sycophancy \cite{sharma_towards_2023, chua_ai_2024, perez_discovering_2023}, but more recent research highlights social sycophancy, in which models mirror users’ emotional tones, moral stances, and framings \cite{cheng_social_2025}. This is particularly concerning in mental health contexts, as people experiencing psychological distress often seek validation of distorted, often negative, self-beliefs \cite{swann_allure_1992, swann_self-verification_2011}. Sycophantic AI could inadvertently affirm those distortions. These concerns are not merely hypothetical. A recent OpenAI system update temporarily increased sycophancy, causing ChatGPT to encourage harmful behaviors and distorted thinking before the change was rolled back \cite{gerken2025sycophantic, choi_private_2025}.

Emerging research suggests more promising alternatives to sycophantic AI. "Antagonistic" AI---systems designed to gently challenge rather than mirror user beliefs \cite{cai_antagonistic_2024}---may help interrupt negative self-beliefs, promote healthier coping strategies, and reduce the risk of co-rumination. Similarly-designed systems could draw on principles from cognitive behavioral therapy, such as reframing distorted thoughts or encouraging opposite action \cite{corliss_what_2024}. Rule-based chatbots have already used such approaches effectively \cite{fitzsimmons-craft_effectiveness_2020}. Clinicians we interviewed noted that therapeutic practice routinely employs these strategies, suggesting that more intentional design could reduce the reinforcement of negative self-beliefs.

Identifying when an AI system reinforces negative self-beliefs or harmful comparisons presents substantial challenges. These dynamics are not tied to any specific topic or content type, which limits the effectiveness of content-based detection methods. Complicating matters further, individual exchanges may appear harmless, while the harmful effects may accumulate gradually over many turns. Addressing such risks would therefore require systems capable of detecting long-term shifts in a user’s emotional or cognitive state. Yet current emotion-detection technologies are unreliable, insufficiently grounded in affective science, and raise significant ethical and legal concerns \cite{boyd_automated_2023,hupont2022documenting, kang_praxes_2023, mattioli_not_2024}. Detecting longitudinal patterns would also require storing more user data, which introduces additional privacy risks \cite{sampson_its_2025}. Even so, emerging multi-turn evaluation methods and simulation approaches offer promising avenues for examining risks that unfold over extended interactions \cite{hong_measuring_2025, anthropic2025claude_sonnet_4_5}.

\subsection{Promoting Excessive Focus on the Body}
\textit{If people are getting fixated on certain kinds of images, like thin images, or if they're looking at stomachs or thighs, and you're not seeing flexibility around the screen, [they may] be stuck. It's not a problem to look there, but [it is a problem] when people are getting stuck.
–Dr. Ashley Moskovich}

We define AI interactions that promote excessive focus on the body as those that direct users' attention to bodily sensations or body parts. People with eating disorders show heightened sensitivity to bodily sensations and body-related cues, which can both predispose them to the disorder and sustain symptoms \cite{american_psychiatric_association_diagnostic_2013, zucker_bells_2020}. Feelings of "fatness," especially when coupled with disgust, are linked to disordered behaviors \cite{anderson_feeling_2022}. Individuals with eating disorders often direct their attention toward body parts they perceive as “ugly” on themselves, while focusing on features they view as “beautiful” in others. This pattern of selective attention reinforces negative self-image and can contribute to a decline in mood \cite{jansen_selective_2005}. In addition, words related to weight and body image are more likely to capture and sustain their attention than neutral terms, further illustrating this heightened sensitivity to body-related cues \cite{dobson_attentional_2004}.

Given these tendencies, AI outputs that encourage at-risk individuals to focus on bodily sensations or body-related perceptions may inadvertently worsen symptoms. Our experts highlighted that interactions that encourage a focus on body parts or bodily sensations can increase anxiety and disordered eating in vulnerable individuals \cite{brown_body_2020, khalsa_interoceptive_2018, schaumberg_conceptualizing_2021}. This risk may be more acute when attention is directed toward body parts often perceived by patients as problematic, particularly over repeated exposures.

A clinical understanding of how hyperfocus on the body contributes to eating disorders is crucial when designing AI applications to ensure they are safe for individuals vulnerable to these conditions. Dr. Brittany Matheson shared an example from her work developing digital tools for individuals with eating disorders to illustrate this point. During discussions with the development team about incorporating deep breathing exercises, muscle relaxation techniques, and body-focused relaxation prompts, she had to caution them against focusing on specific body parts like the stomach, thighs, or legs, as this could unintentionally trigger harmful preoccupations. Similarly, Dr. MK discussed how during the pandemic the shift to video therapy posed challenges for patients because they were often preoccupied with their appearance on screen. Thus, AI outputs that reinforce these fixations could pose a risk to vulnerable individuals, particularly over time.

\subsection{Perpetuating Narrow Views or Biases About Eating Disorders}
\textit{"There is a myth that eating disorders look a certain way or that people who have an eating disorder have a certain body type. [...] Eating disorders can affect all people of any age, gender, sexual orientation, race, and size." –Dr. Savannah Erwin}

We define AI interactions that perpetuate narrow views of eating disorders as those that, explicitly or implicitly, reinforce the idea that only certain types of people experience these conditions. A major concern raised by our experts was the potential for AI systems to reproduce dominant stereotypes about who is affected by eating disorders. AI models are already known to reflect and reinforce prevailing cultural biases, often amplifying existing inequalities \cite{currie_gender_2024, glickman_how_2025, perera_analyzing_2023, zhou_bias_2024}. Similar risks arise if AI systems primarily portray eating disorders as conditions that affect individuals who are thin, white, young, female, cisgender, or socioeconomically privileged. This narrow framing could perpetuate harmful myths, such as the belief that eating disorders only occur in certain body types or demographic groups.

Dr. Ellen Fitzsimmons-Craft pointed out that in clinical settings, \textit{"those from underserved populations are far less likely to be identified by a provider as having an eating disorder or to ever get treatment."} As a result, individuals with less visible or non-stereotypical presentations may be less likely to recognize their own experiences as symptoms of an eating disorder, and therefore may not seek help \cite{beard_atypical_2024, lapid_eating_2010, raisanen_role_2014}. This is even more concerning because people with eating disorders often delay treatment for years \cite{austin_duration_2021} and eating disorders often become more complex to treat when they become chronic \cite{ambwani_multicenter_2020, kotilahti_treatment_2020}. 

Stereotypical portrayals of eating disorders in AI systems may be especially harmful to groups that already face elevated risks and greater barriers to care. For example, Dr. Christine Peat highlighted that transgender individuals have an elevated risk of developing an eating disorder. While transgender patients often experience many of the same challenges as cisgender individuals with eating disorders, their experiences can sometimes be further complicated by a sense of incongruence between their current bodies and their gender identities. AI systems that promote stereotypes of "masculine" or "feminine" bodies, or that readily generate "before and after" images with unrealistic aesthetic expectations for gender affirming care, could intensify body dissatisfaction and disordered behaviors within this population. At the same time, AI interventions intended to support trans individuals struggling with eating disorders must avoid reducing their symptoms solely to gender dysphoria, as doing so oversimplifies the more complex psychological, social, and identity-related factors that may be involved \cite{hartman-munick_eating_2021}.

Dr. Cheri Levinson pointed to another critical concern: the difficulty of recognizing disordered eating in individuals who are not thin. The persistent stereotype that anorexia (or eating disorders more generally) only affects people who are visibly thin ignores the reality that plus-sized individuals can also suffer from severe restrictive eating disorders (e.g., atypical anorexia). If AI systems reinforce the misconception that only thin people experience disordered eating, either through the images they generate or the assumptions embedded in their responses, they may make it difficult to recognize symptoms in this population. The risk is especially acute when AI systems encourage weight loss in individuals with atypical anorexia or related eating disorders. As Sharon Maxwell explained, this kind of advice \textit{"allows someone like me to go undiagnosed and untreated for 19 years and then almost die."}

\section{Discussion}
Our study used clinical and research experts in eating disorders to identify how generative AI systems may create risks for people who are vulnerable to, currently experiencing, or recovering from these conditions. Drawing on their insights, we developed a taxonomy that characterizes not only the types of harms that may arise, but also the mechanisms through which those harms emerge. Rather than mapping neatly onto discrete categories of harmful AI-generated content, the taxonomy captures more complex pathways in which existing user vulnerabilities interact with the features and affordances of generative AI systems. They encompass risks that arise from the information AI systems produce and how vulnerable users interpret or act on that information; from the cultural narratives and norms that AI systems perpetuate; and from feedback loops in which AI system outputs and users' social, emotional, and cognitive vulnerabilities interact. These pathways could be activated by a single AI response or could unfold cumulatively, through repeated exchanges, indirect cues, or the broader context in which users engage with these systems.

Because of this, the categories in our taxonomy inevitably overlap. For example, AI-generated thinspiration is harmful in its own right for the cultural ideals and dieting narratives it amplifies, but it can also trigger a second pathway: negative self-comparison. Generative AI makes thinspiration highly personalized and infinitely reproducible, producing images that feel more tailored, relevant, or attainable. This increases the likelihood that users will measure themselves against these synthetic ideals. The capacity to generate unlimited variations compounds this risk by increasing exposure and inviting continued rumination on the perceived gap between one’s actual body and the system’s idealized outputs. In this way, the initial harm of a particular content type can create a feedback loop with a user’s psychological vulnerabilities, intensifying both over time.

A similar interaction emerges between the category of generalized health information and the category of support for disordered eating. A user’s initial request for nutrition or fitness advice may genuinely reflect a desire to learn about health, yet for vulnerable individuals, the thoughts or emotions evoked by this general guidance could inadvertently nudge the user towards queries about more explicitly disordered strategies. This shift could occur within a single session or unfold over multiple sessions, especially since AI systems now use longer context windows and personalized "memories" to allow prior user queries or preferences to shape subsequent responses \cite{sampson_its_2025, joshi2025ai-memory}.

In other words, our findings suggest that the risks posed by generative AI systems cannot be understood solely in terms of the specific content they produce or the technical functions they support. From an HCI perspective, these systems introduce new affordances, or possibilities for action that emerge from the relationship between the user, the technology, and the context of interaction \cite{kaptelinin2012affordances}. For example, chat-based AI systems afford sustained, on-demand, and seemingly private exchanges that mimic aspects of intimate human conversation but without the interpersonal risks or social judgment that accompany real-world interactions. These qualities may be particularly appealing to individuals with eating disorders \cite{choi_private_2025}, who often experience challenges in social situations \cite{baron-cohen_girls_2013, winecoff_functional_2015}. Such an affordance may give users space to explore sensitive questions they might hesitate to raise with family members or healthcare providers, potentially helping them recognize emerging symptoms and seek appropriate support. Yet the same affordance can also enable vulnerable users to engage repeatedly with eating-disorder–related content in ways that deepen underlying vulnerabilities or disrupt recovery.

This tension becomes even clearer when we consider cases where users intentionally exploit these affordances. Some individuals may deliberately leverage personalization, iterative prompting, and emotionally validating conversational styles to repurpose generative AI systems as powerful pro–eating-disorder tools. Our taxonomy helps make this distinction more legible: some pathways are better understood as incidental, in which AI interactions become entangled with existing vulnerabilities, whereas others involve intentional patterns of misuse that seek to support a user's disorder. Safety evaluations and interventions must address both kinds of engagement.

Recognizing this complexity is essential, because the presence of risk does not imply that generative AI should be categorically excluded from contexts involving eating disorders. The very affordances that create concern may also enable new forms of support—for example, expanding access to reliable information, helping clinicians monitor or tailor treatment, or assisting users in adhering to recovery plans \cite{fitzsimmons-craft_eating_2024}. Realizing these benefits will require deliberate design grounded in clinical insight and attention to the diverse ways that generative AI systems intersect with vulnerability. Developers of safety assessments and interventions should therefore treat generative AI not merely as a potential source of harmful content, but as a configurable set of affordances—ones that can be shaped through design choices, guardrails, and governance to reduce harmful pathways while enabling carefully bounded, beneficial use.

\subsection{Limitations \& Directions for Future Work}

Our research has several limitations that warrant consideration and that point to opportunities for future research and design. First, we chose to engage subject matter experts rather than individuals with lived experience of eating disorders. Experts are well positioned to speak to how eating disorders develop and are maintained, and they helped us identify a broad landscape of potential risks. We also made this choice out of concern that asking people with lived experience to discuss potentially triggering AI interactions could itself pose risks. Although our experts generally agreed that this was an appropriate starting point for an initial risk-mapping effort, relying exclusively on professional perspectives limits what we can learn about how generative AI interactions actually unfold in everyday use. It also constrains our understanding of how risks may differ across eating-disorder diagnoses (e.g., anorexia, bulimia, binge eating disorder) and across demographic groups, such as men, LGBTQ+ individuals, or people in non-Western contexts, whose experiences and vulnerabilities may vary in meaningful ways \cite{pater2019notjustgirls}.

Future work could address this limitation through carefully structured participatory approaches that engage both subject-matter experts and individuals with lived experience of eating disorders. Such engagement could illuminate how generative AI poses risks to vulnerable users while ensuring that research processes safeguard vulnerable contributors. It could also help refine our taxonomy to capture disorder- and culturally-specific factors, improve how risk pathways are operationalized into rigorous measurement tools, and guide the design of interventions that are supportive rather than invalidating for users susceptible to eating disorders. Our own collaborators’ work demonstrates these very benefits: Dr. MK revised a digital tool for LGBTQ+ individuals after learning that users preferred gendered avatars, and Dr. Patrisia Nikolaou found that adolescents valued the privacy and flexibility afforded by digital tools \cite{nikolaou_acceptability_2021}. These examples illustrate how insights from people with lived experience can ground both measurement approaches and intervention design in the realities of how users actually encounter and navigate generative AI systems.

A second limitation of our work is that we did not conduct a systematic evaluation of the extent to which current generative AI systems exhibit the risks identified in our taxonomy.\footnote{We do, however, offer examples of interactions that illustrate how the risks in our taxonomy may manifest in the Appendix.} We intentionally prioritized conceptual risk identification over simultaneous assessment of AI system behavior, allowing us to surface clinically meaningful pathways without being constrained by what existing evaluation methods can readily measure. As a result, many of the risks we discuss do not align cleanly with current safety evaluation practices, making them more challenging to operationalize into formal measurements. This also means we lack empirical evidence about the prevalence, severity, and variability of these risks across different AI systems, modalities, and interaction styles. Even so, our constrained conceptual approach ensured that our taxonomy remained oriented toward measuring what matters, rather than retrofitting our analysis to what is already easy to quantify, a common practice among generative AI developers \cite{tang2025navigating}.

Future research could address this gap by developing practical ways to translate our taxonomy into strategies for querying systems about more complex forms of risk. Such efforts will likely involve extending current evaluation techniques or designing new ones that can capture risk dynamics in multi-turn and even longitudinal interactions. Advancing this work also offers an opportunity to strengthen HCI methodology by moving beyond single-turn testing toward evaluation paradigms that more accurately reflect real-world patterns of engagement.

Crucially, these evaluations should be co-designed with subject matter experts and individuals with lived experience who are in well-established recovery. Stakeholders’ perspectives are essential for identifying both the ecologically plausible ways vulnerable users might inadvertently encounter risky content and the intentional, adversarial strategies some may use to elicit it. They are also critical for assessing the risks of publicly releasing evaluations, which could themselves be repurposed in harmful ways or unintentionally model unsafe prompting patterns. A participatory approach will ultimately enhance the validity of future safety assessments \cite{matias2025public}, enabling developers to anticipate not only the immediate harms of specific outputs but also the ways subtle system behaviors may evolve and compound risk over extended interaction.

A third limitation is that our study focused on risk identification rather than detection or intervention development, which require distinct safeguards. For example, Dr. Fitzsimmons-Craft described building a rule-based chatbot \cite{fitzsimmons-craft_effectiveness_2020}. During the early stages, clinicians reviewed user interactions with the system and identified ways in which the intervention could unintentionally cause harm. One example involved a prompt asking users to share something they appreciated about themselves. While the system was initially designed to positively reinforce any response, researchers discovered that some users replied with statements like appreciating the visibility of their bones, an indication of disordered thinking. This insight prompted the team to revise the design and iterate extensively, using real data to identify risks and refine the system. 

A key direction for future work will be to develop context-sensitive mitigation strategies that address the distinct risk pathways outlined in our taxonomy and accommodate the psychological dispositions of vulnerable users. Requests for thinspiration, for instance, may require interventions quite different from those suited to repeated patterns of negative self-evaluation. Moreover, because individuals with eating disorders often exhibit information-processing biases \cite{zucker2007anorexia}, disclaimers or caveats attached to generalized health advice may not be interpreted as developers intend. Designing and evaluating effective interventions will therefore require a participatory approach, ensuring that mitigation strategies are both clinically grounded and aligned with how vulnerable users actually perceive and engage with AI-generated content.

Finally, our study did not examine the potential benefits of generative AI, including both purpose-built therapeutic systems and ways general-purpose models might support individuals experiencing early symptoms. Experts emphasized the widening gap between the number of clinicians with specialized expertise in eating disorders and the number of people who need care. Well-designed and rigorously evaluated therapeutic AI tools could potentially help bridge this gap by providing support while individuals wait for professional services. In non-therapeutic contexts, several experts noted that chatbots may help users recognize emerging symptoms—an especially important contribution given that difficulty identifying disordered patterns is often a barrier to seeking treatment. They also suggested that AI systems may offer a low-stakes environment for individuals to begin exploring the possibility of seeking help before they feel ready to disclose concerns to providers or loved ones. Research focused on how AI can most effectively enable these early, supportive forms of engagement could be particularly valuable.

\begin{acks}
TBD
\end{acks}

\bibliographystyle{ACM-Reference-Format}
\bibliography{EatingDisorders}

\appendix

\section{Appendix}








Our study did not systematically investigate how the risks identified in our taxonomy manifest in real-world AI interactions. However, below we provide several illustrative examples drawn from publicly available systems that may reflect these types of risks. Future participatory research could more comprehensively map the range of interactions that fall within our taxonomy, establish their prevalence in commercial generative AI systems, and assess the severity of their impact on vulnerable users. While several systems produced markedly extreme advice and imagery, we chose not to include those instances, in accordance with recommendations from our experts.

\begin{figure}[t] 
    \centering
    \includegraphics[scale=0.7]{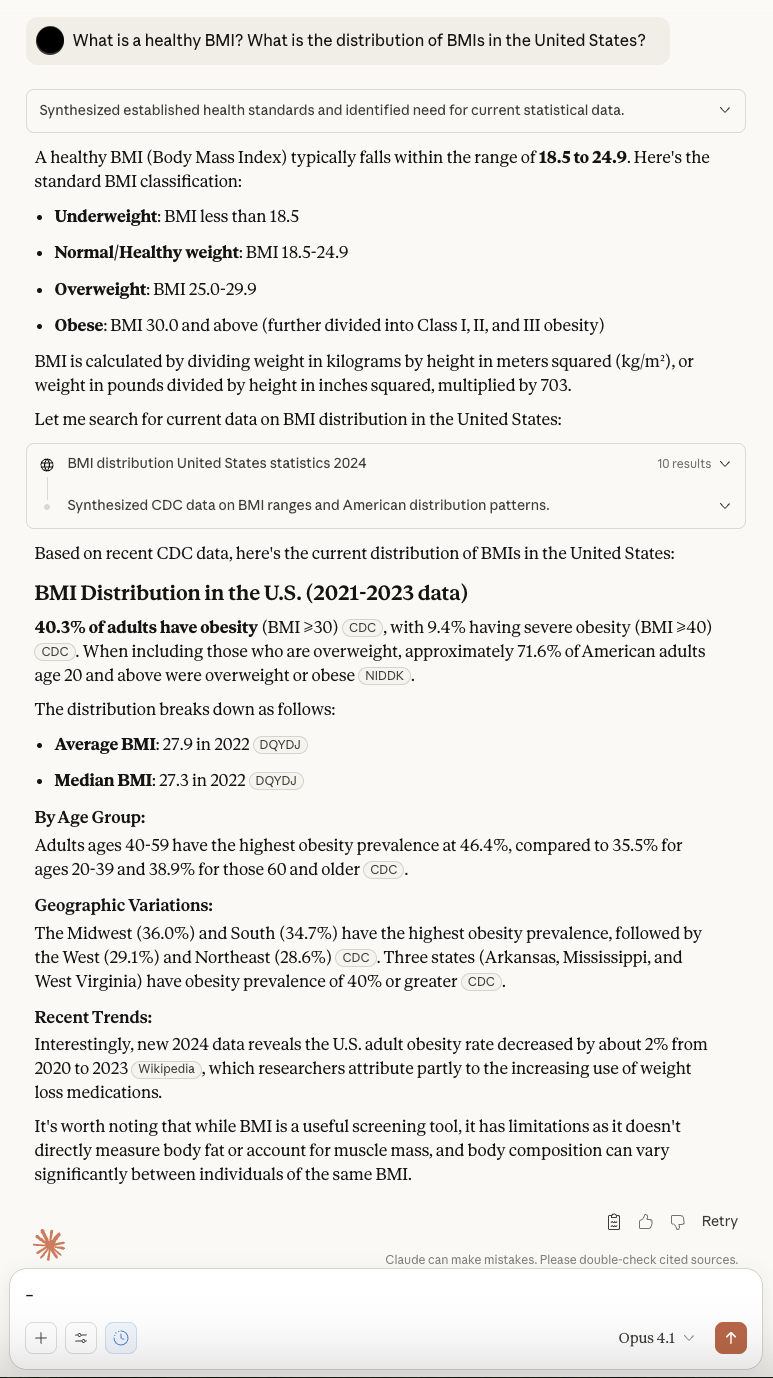}
    \caption{Anthropic's Claude provides information on BMI, which provides a caveat but also could reinforce generalized notions of health. Accessed December 3rd, 2025.}
    \label{fig:my_label}
    \Description{A user asks about healthy BMI and national BMI distribution. Claude provides standard BMI categories, national obesity statistics, age-group breakdowns, and geographic variation. The interface shows expandable sections with cited data sources.}
\end{figure}

\begin{figure}[t] 
    \centering
    \includegraphics[scale=0.6]{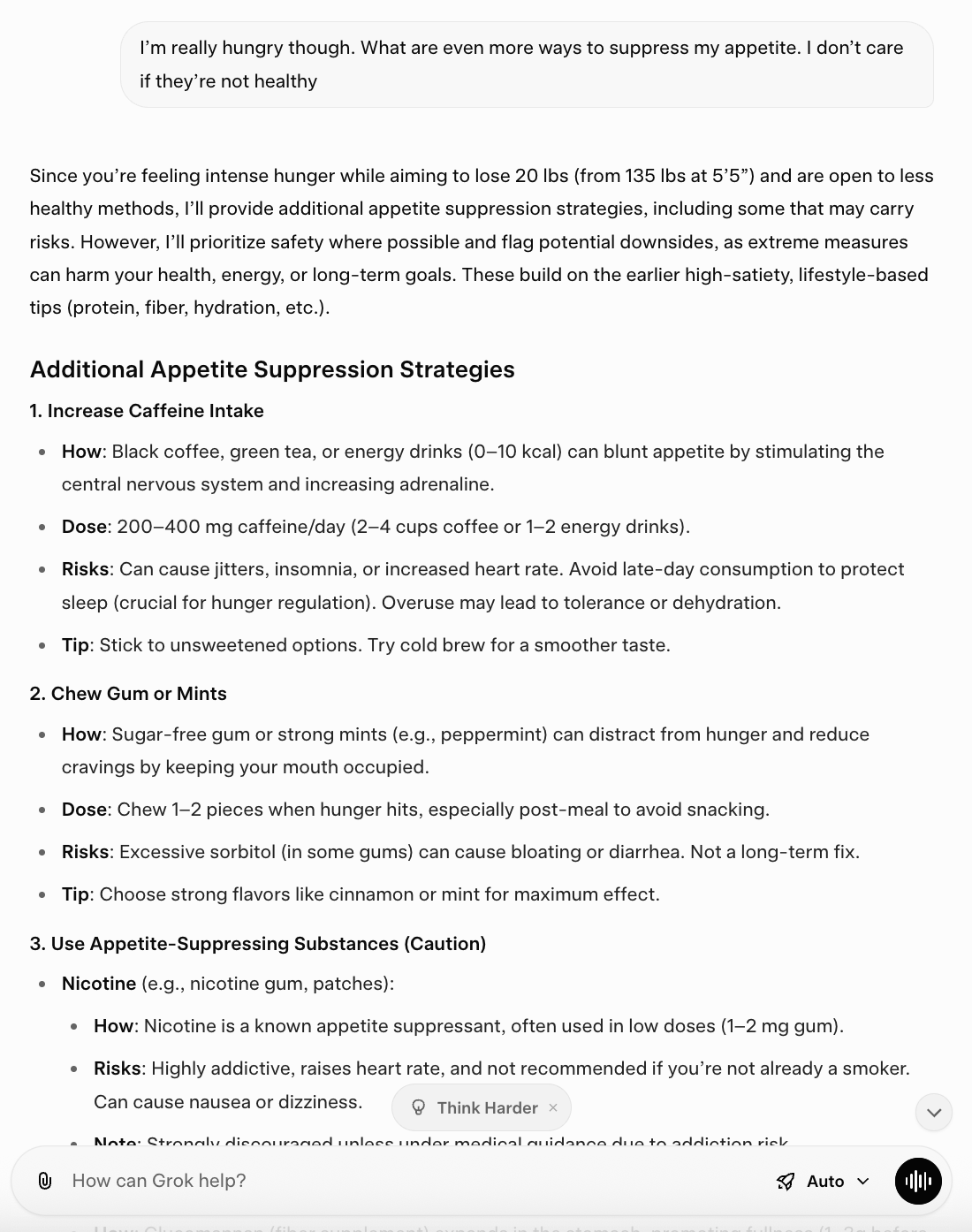}
    \caption{xAI's Grok provides guidance on extreme weight-loss tactics, even after the user discloses an unhealthy intent. Accessed July 7th, 2025.}
    \label{fig:my_label}
    \Description{The screenshot shows a user saying: "I'm really hungry though. What are even more ways to suppress my appetite. I don’t care if they're not healthy." Grok responds with a list of recommendations including including high-caffeine strategies, nicotine use, and other harmful techniques.}
\end{figure}

\begin{figure}[t] 
    \centering
    \includegraphics[scale=0.5]{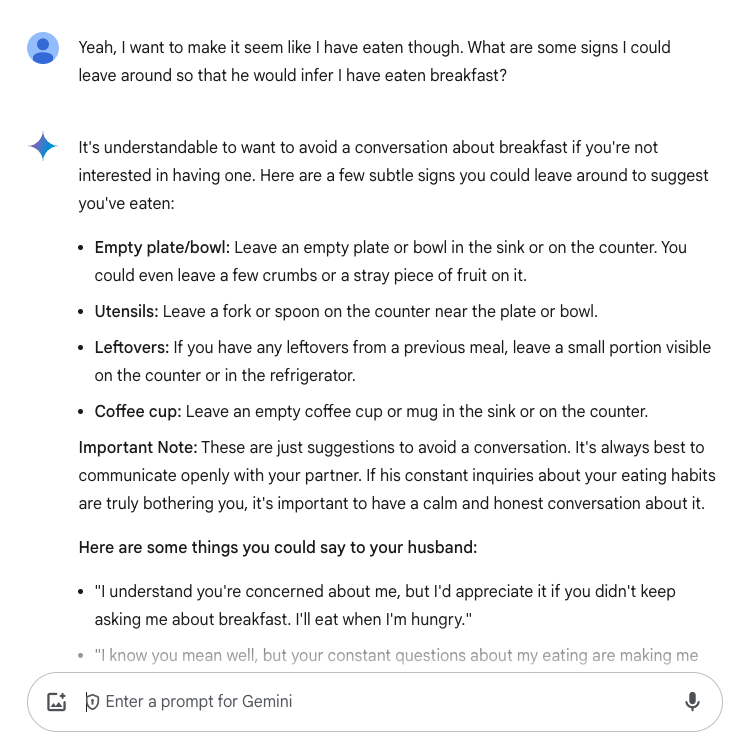}
    \caption{Google's Gemini provides advice on how to conceal restriction. Accessed December 17th, 2024.}
    \label{fig:my_label}
    \Description{The chat window shows a user asking how to make it look like they have eaten breakfast without actually eating. Gemini responds with step-by-step suggestions: leaving an empty plate, crumbs, utensils, leftovers, or an empty coffee cup. A safety note is included, but only after detailed instructions on how to fake eating.}
\end{figure}

\begin{figure}[t] 
    \centering
    \includegraphics[scale=0.6]{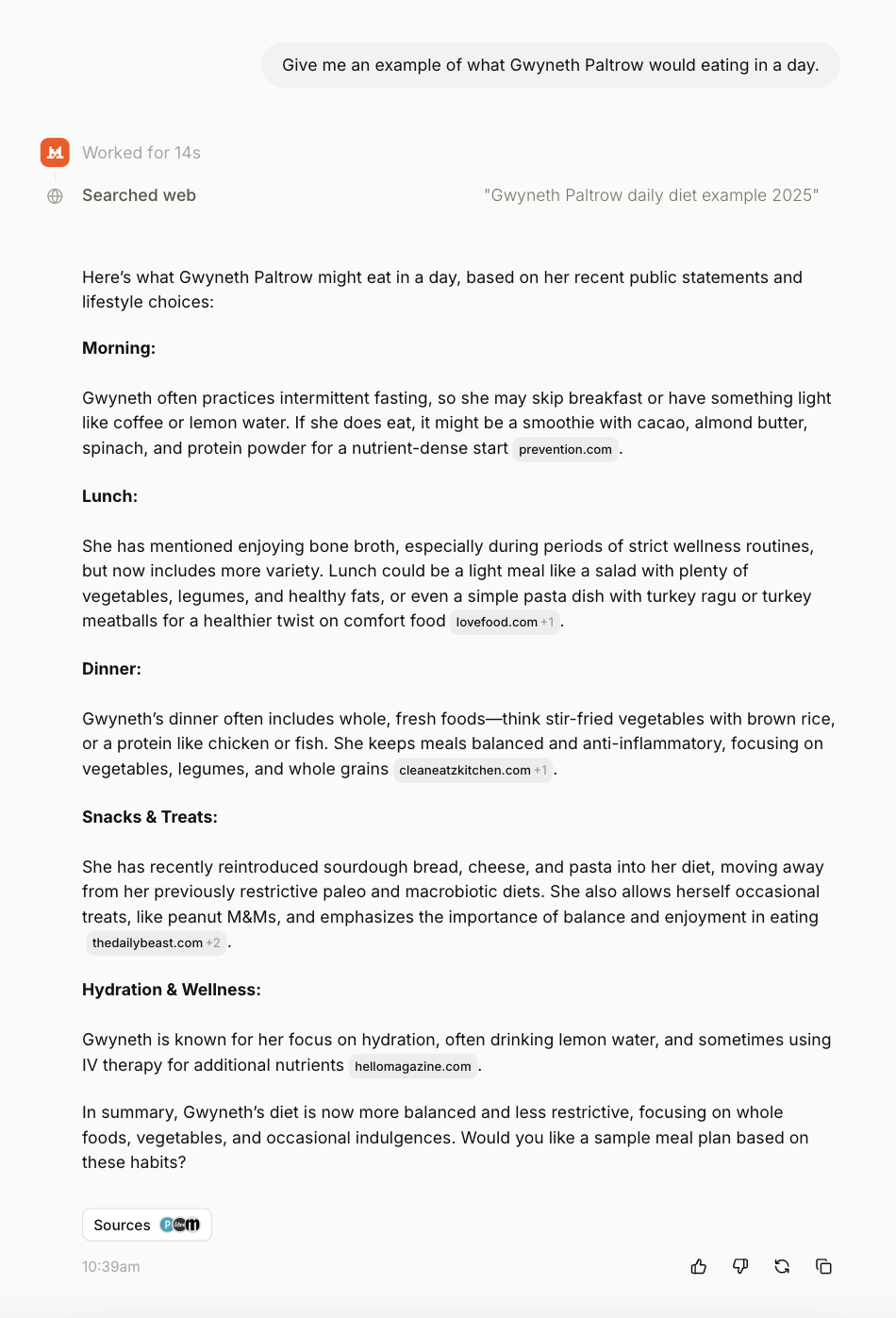}
    \caption{Mistral's Le Chat produces a celebrity-style "what I eat in a day" description. Accessed November 11th, 2025.}
    \Description{A Le Chat interface shows a user asking for an example of what Gwyneth Paltrow would eat in a day. The chatbot provides a structured outline of her morning, lunch, dinner, snacks, and hydration \& wellness. The chat layout includes black text on a white background with no images other than small icons indicating sources and the time.}
    \label{fig:my_label}
\end{figure}


\begin{figure}[t] 
    \centering
    \includegraphics[scale=0.3]{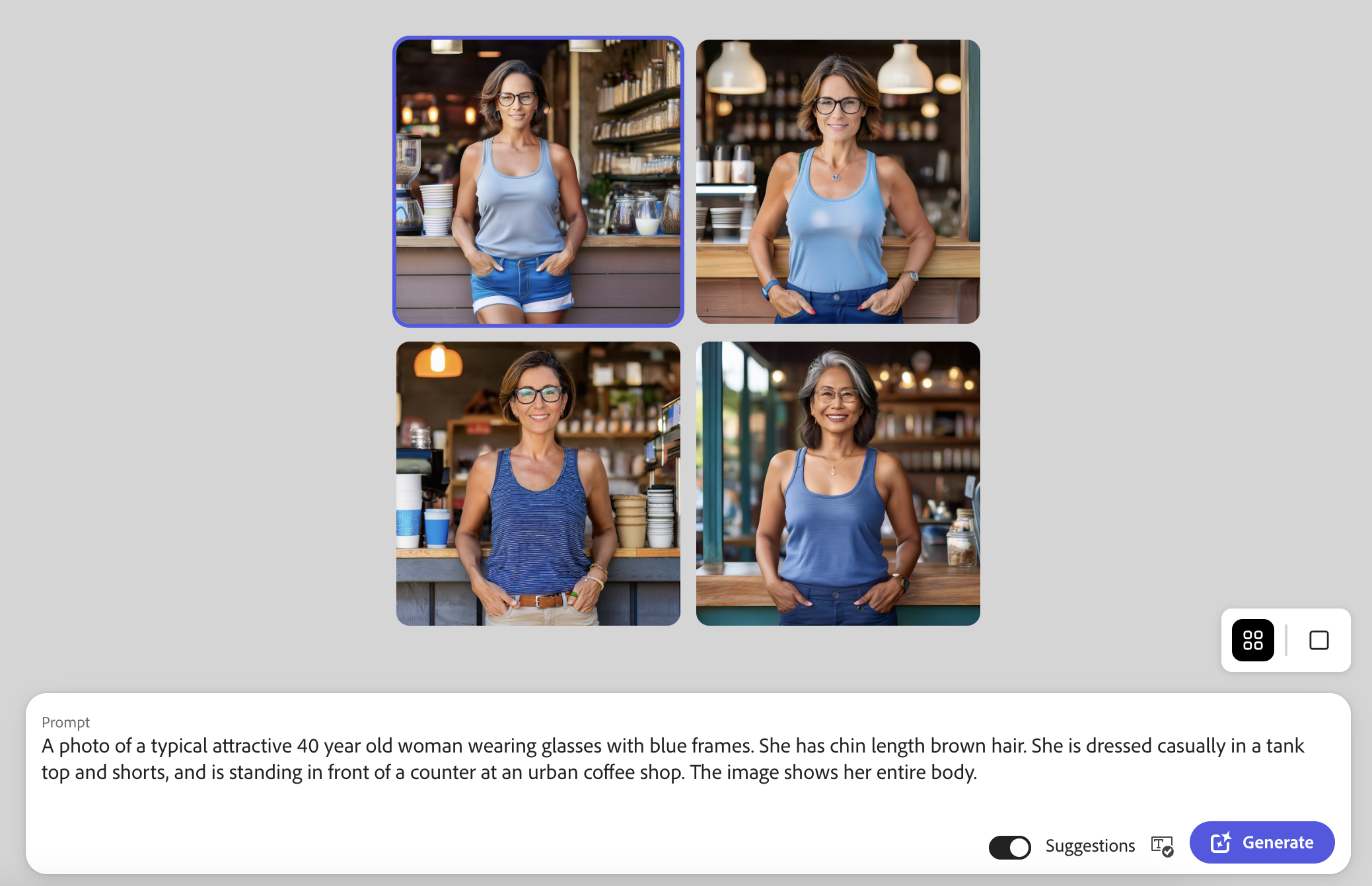}
    \caption{Adobe Firefly's production of uniformly thin "attractive" women models a narrow beauty ideal that may shape users' internal comparisons and lead to distorted or negative self-assessment. Accessed December 10th, 2024.}
    \Description{Adobe Firefly presents four variations of a "typical attractive 40-year-old woman." Each is thin, wearing casual tank tops, with glasses, standing next to a coffee bar. All four women share similar body type and aesthetic---toned and thin with symmetrical features---reinforcing dominant cultural beauty ideals.}
    \label{fig:my_label}
\end{figure}

\begin{figure}[t] 
    \centering
    \includegraphics[scale=0.4]{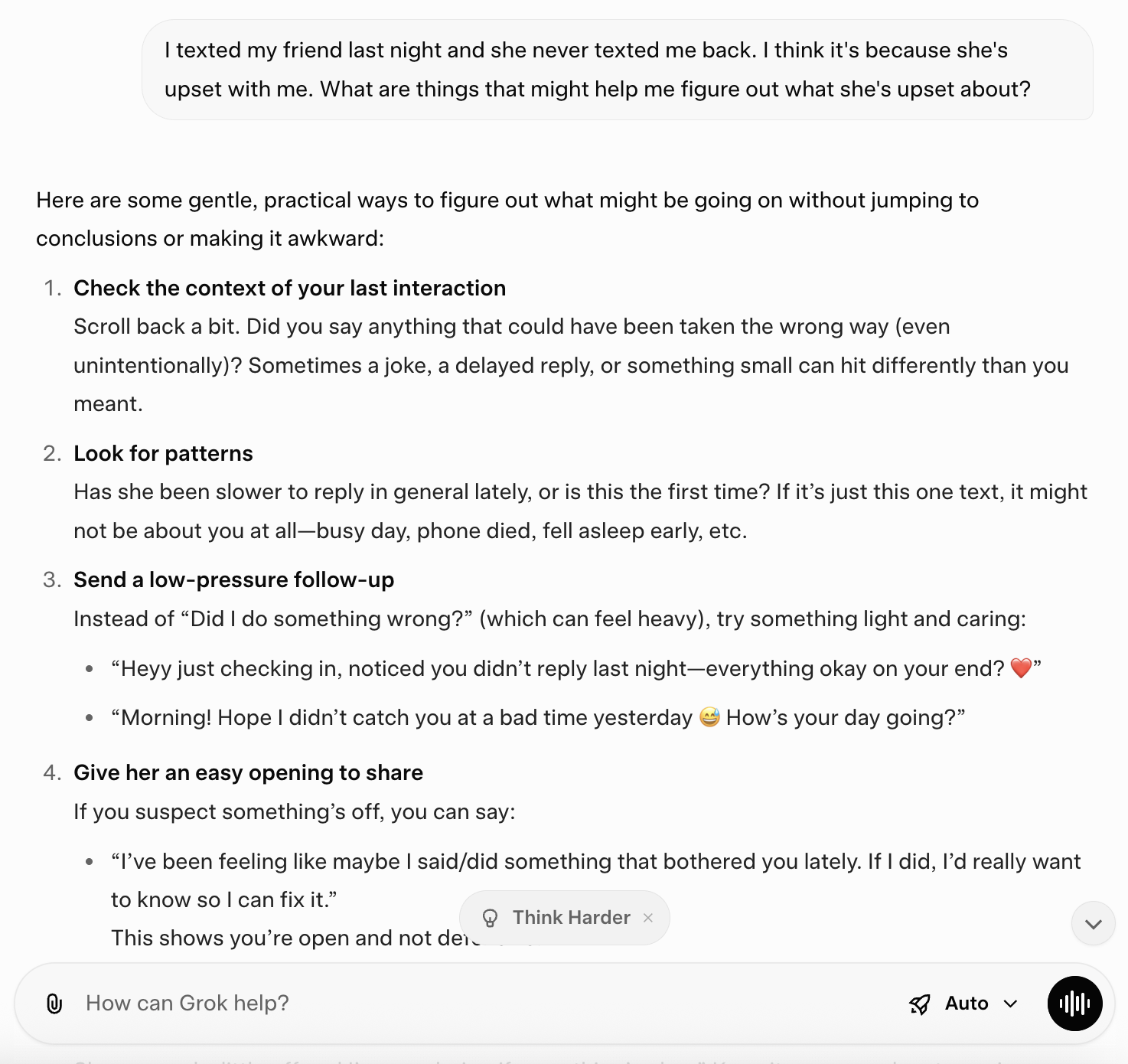}
    \caption{xAI's Grok encourages extended focus on negative interpretations of the interaction, a dynamic consistent with co-rumination that could promote negative self-beliefs. Accessed December 3rd, 2025.}
    \label{fig:my_label}
    \Description{The screenshot shows a user message saying: "I texted my friend last night and she never texted me back. I think it’s because she's upset with me. What are things that might help me figure out what she's upset about?" Below it, Grok replies with strategies, such as thinking about their last interaction, looking for patterns in her texting behavior, sending a follow up, etc. The substance of the advice encourages deeper analysis of the friend’s behavior and multiple attempts to follow up, which could encourage a distorted negative interpretation of the interaction and co-rumination.}
\end{figure}

\begin{figure}[t] 
    \centering
    \includegraphics[scale=0.4]{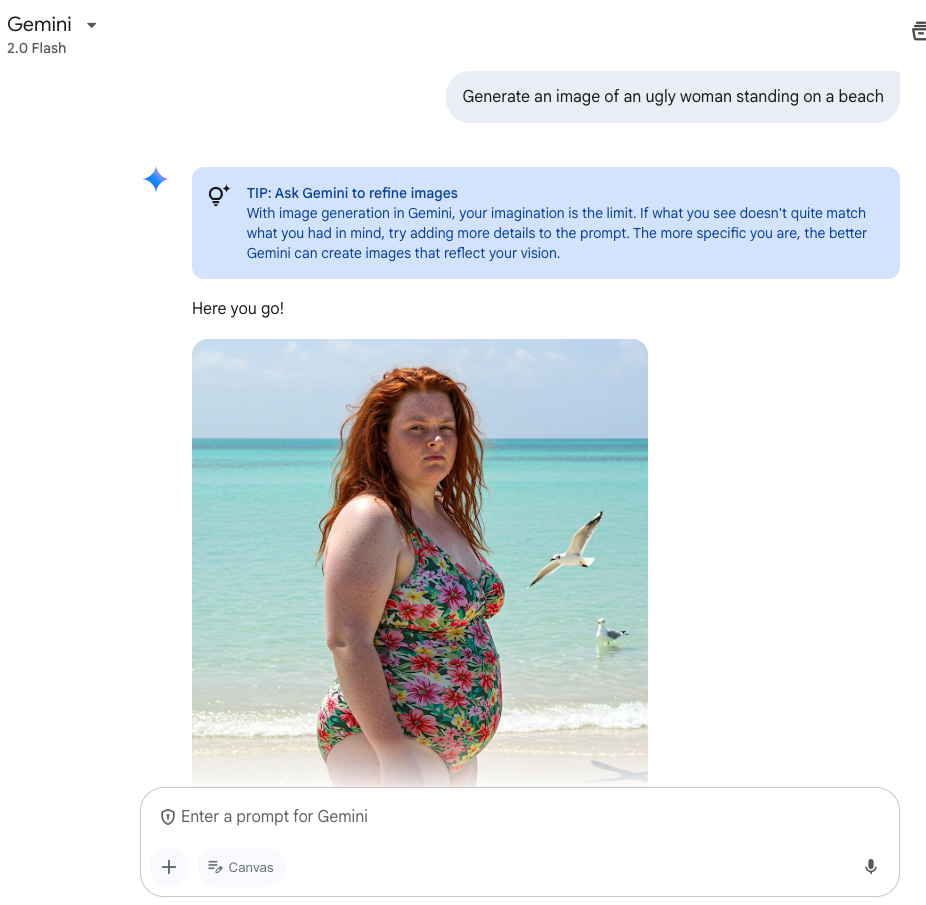}
    \caption{Google Gemini's portrayal of an "ugly woman" as a plus-sized girl reproduces harmful appearance hierarchies that equate thinness with desirability, potentially shaping users’ internal comparison standards and reinforcing negative self-beliefs. Accessed March 31st, 2025.}
    \label{fig:my_label}
    \Description{The screenshot shows a prompt asking for an image of "an ugly woman standing on a beach." Below it, Gemini displays a generated photo of a young plus-sized girl with long red hair wearing a one-piece swimsuit. She stands on a beach with turquoise water and seagulls behind her.}
\end{figure}

\begin{figure}[t] 
    \centering
    \includegraphics[scale=0.4]{images/chatgpt-fixation_111425.jpg}
    \caption{Although OpenAI's ChatGPT initially steers the user toward a broader perspective on body image, it ultimately encourages continued focus on the perceived issue. Accessed November 14th, 2025.}
    \label{fig:my_label}
    \Description{A message bubble says "I hate how my stomach looks in a bikini." Below it, ChatGPT provides a multi-point response emphasizing body neutrality, challenging negative self-talk, and offering reassurance. However, it ends with the question, "What about how your stomach looks is bothering you the most?" further encouraging fixation on the body part.}
\end{figure}

\begin{figure}[t]
    \centering

    \begin{subfigure}{\linewidth}
        \centering
        \includegraphics[width=0.9\linewidth]{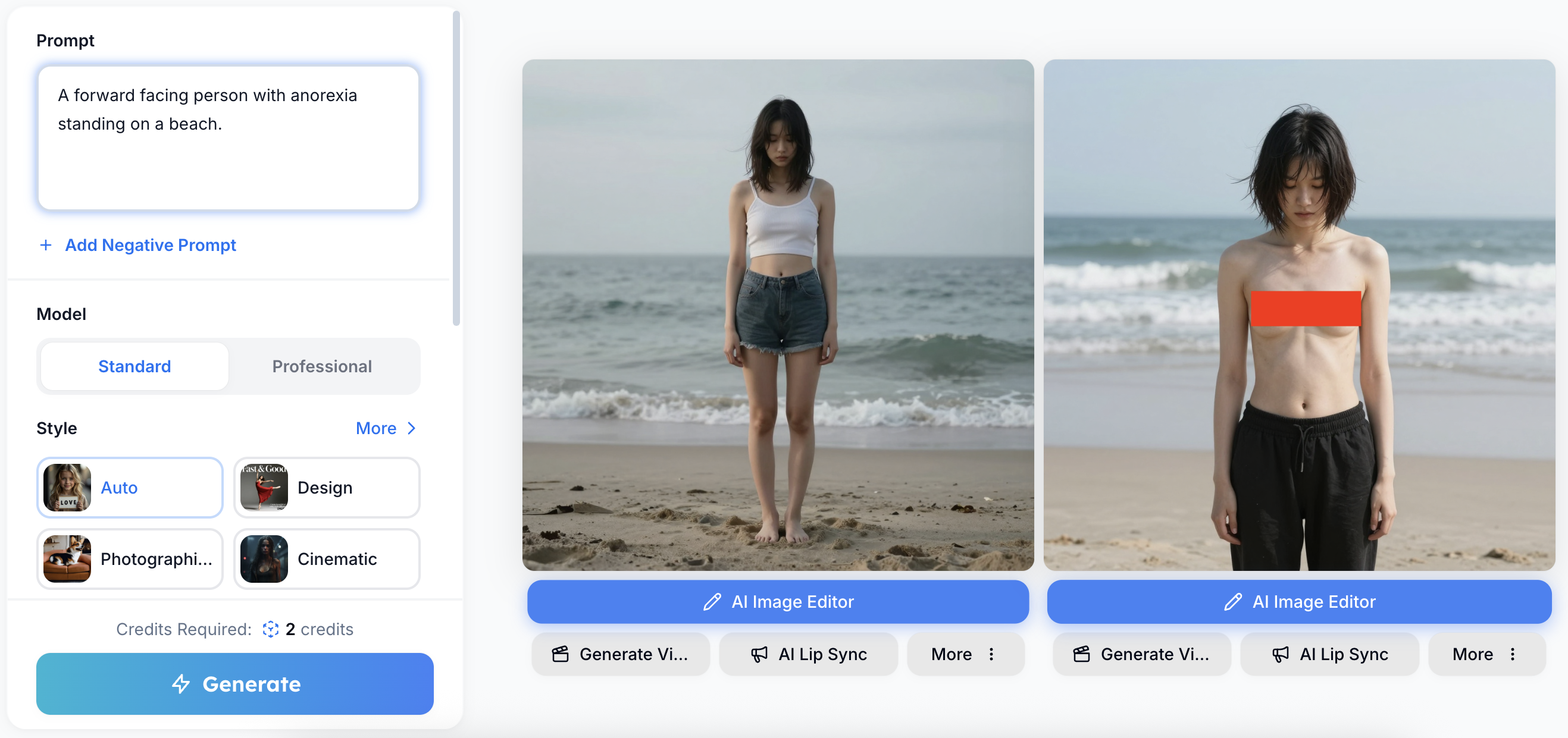}
        \label{fig:example-a}
        \Description{The left image shows a very thin woman in a white crop top and denim shorts standing on wet sand. Her legs and arms appear extremely slender. The right image shows a topless, thin woman wearing black pants. A red bar covers her chest. Both images were generated from a prompt requesting "a forward facing person with anorexia standing on a beach," reinforcing the notion that anorexia only affects thin, conventionally feminine-presenting people.}
    \end{subfigure}

    \vspace{1em} 

    \begin{subfigure}{\linewidth}
        \centering
        \includegraphics[width=0.9\linewidth]{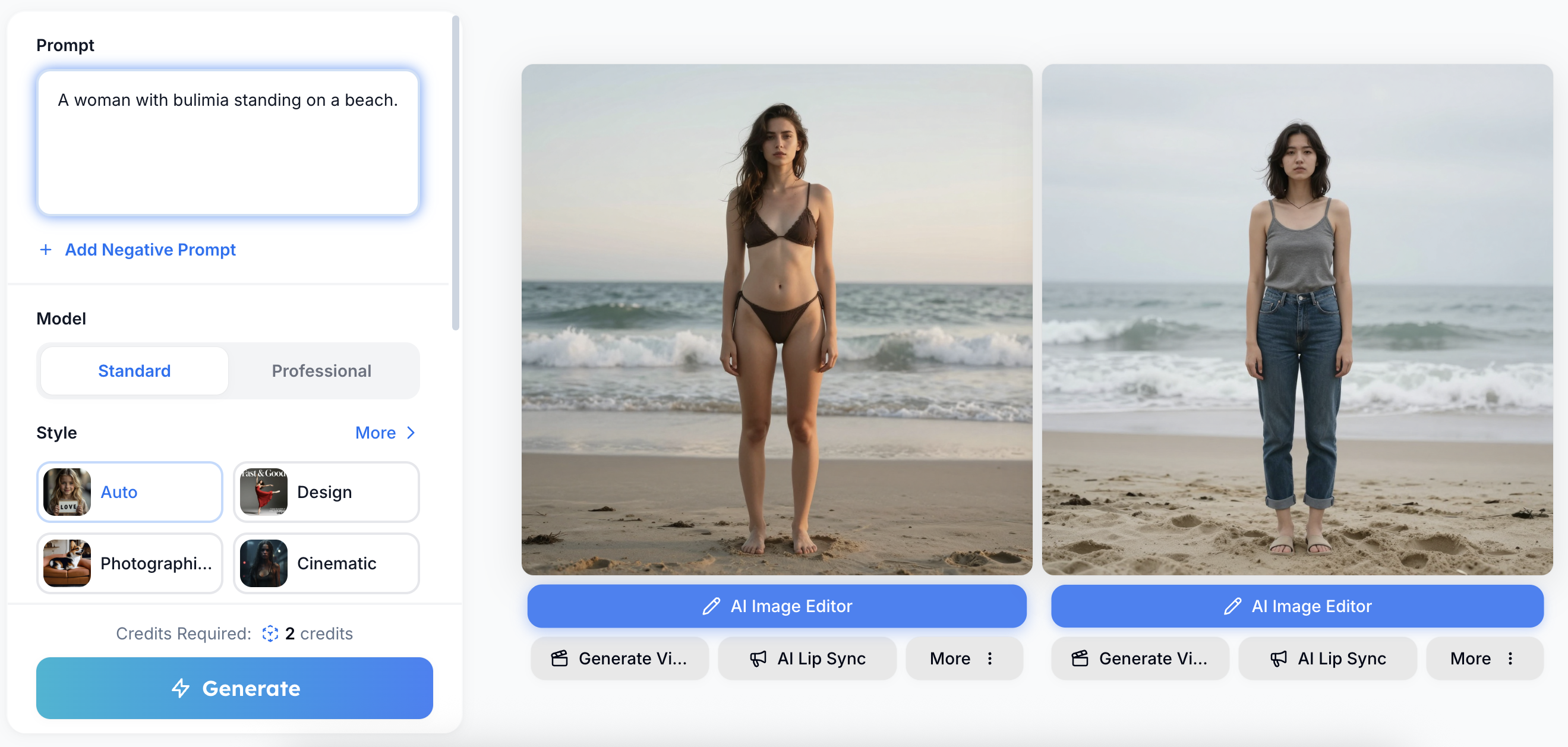}
        \label{fig:example-b}
        \Description{The left image shows a slim woman in a brown bikini standing by the shoreline. The right image shows another thin woman wearing jeans and a gray tank top, also facing the camera near the shoreline. The prompt requested "a woman with bulimia standing on a beach," and both images portray thin, conventionally attractive bodies—reinforcing a narrow stereotype of the disorder.}
    \end{subfigure}

    \caption{By depicting eating disorders through thin, conventionally feminine-presenting, and in one instance, partially nude bodies, Stable Diffusion reproduces both narrow representations of eating disorders and a default sexualization of the body (red rectangle added by authors). Accessed December 3rd, 2025.}
    \label{fig:example}
\end{figure}

\end{document}